\documentclass[6pt]{article}
 \pdfoutput=1
\usepackage{indentfirst}
\usepackage[letterpaper, landscape, margin=0.75in]{geometry}
\usepackage{amsmath,amssymb,amsthm}
\usepackage{hyperref}

\author{Stanislav Srednyak}
\title{Feynman integrals as flat bundles over the complement of Landau varieties.}
\begin{document}
\maketitle

\begin{abstract}

We demonstrate that Feynman integrals of a fixed diagram form a flat vector bundle over the complement of Landau varieties that possesses a connection
\begin{equation}
\frac{\partial}{\partial p_{i,\mu}}f_\beta(p_{i,\mu})=\sum_{\beta'} \sum_k \sum_{I_1,...,I_k} \frac{A^{I_1,...,I_k}_{i,\mu,\beta,\beta'}(p)}{L_{I_1}(p)...L_{I_k}(p)} f_{\beta'}(p)
\end{equation}
where $L_I(p)$ are the Landau polynomials (multidiscriminants). This is the Gauss-Manin connection for the original integral. This result suggests a shift of focus from the integrals to the geometry of the complement of Landau varieties and Riemann-Hilbert data associated with these varieties.
\end{abstract}

There is large interest in physics community in differential equations for Feynman integrals ~\cite{Henn,Weinzierl-Stach,Mastrolia}. These equations are known to be regular ~\cite{Weinzierl-Stach}. The fact that the equations form a connection was suggested in ~\cite{Henn}. We add one more line to these results by pointing out that the connection matrix must be a rational function, denominator of which can be identified with product of Landau polynomials of the diagram. These polynomials are determinants of Koszul complexes ~\cite{GZK} and therefore can be algorithmically computed. The matrices $A^{I_1,...,I_k}_{\i,\mu,\beta,\beta'}(p)$ are also polynomials and are constrained by integrability condition and by regularity at infinity of the space of external momenta ( there is a canonical way to compactify this apriori affine space, that results from the fact that the Landau varieties are embedded in projective duals to torus orbits).

The Feynman integral of a diagram is
\begin{equation}
f(p_{i,a,\mu})=\int \prod_{i,a} D(q_i+p_{i,a}) d^nq
\end{equation}
where $q=(q_{1,\mu_1},...,q_{L,\mu_L})$ is the coordinate on the the space $CP^n=CP^{dL}$ of loop momenta, $q_i$ are linear combinations of loop momenta with integer coefficients (see ~\cite{Weinzierl-Bogner}). We compactify the space of loop momenta to $CP^n$, as is customary ~\cite{Connes-Marcolli}, and use dimensional regularization throughout, which specifies that the integrand vanishes "at infinity" of the loop momenta space. To keep notaton simple, we omitted mass variables, but it should be clear from the discussion that our results tautologically generalize to this enlarged space of parameters. The integration is performed over a relative chain. 

We will aslo use 
\begin{equation}
f_\beta(p)=\partial^\beta f(p)=(\partial /\partial p_{1,1,\mu_1})^{\beta_{1,1,\mu_1}}...(\partial /\partial p_{I,a_I,\mu_I})^{\beta_{I,a_I,\mu_I}}f(p).
\end{equation}
The integrand and the space $P=(p_{i,a})$ possess large symmetry, but it will not play a role in our discussion.

{\bf Theorem } The functions $f_\beta$ form a vector bundle over the complement to the Landau varieties of the diagram and satisfy the connection stated in the appendix.

\begin{proof}[Proof]
The proof is essentially obvious. It follows from the following three basic facts. {\bf Fact 1}. The functions $f$ can have singularities only along Landau varieties. This follows from the fact that singularities can arise only when there is a change in topology of the polar locus of the integrand ( this is a standard fact from the theory of vanishing cycles, see e.g. ~\cite{Sabbah}). {\bf Fact 2} There is Gauss-Manin connection for the hyperfunction $f$. This is true because there is GZK D-module on for generic parameters ~\cite{GZK} and holonomic D-moodules can be pulled back ~\cite{Sturmfels}. See also ~\cite{Weinzierl-Stach}. {\bf Fact 3}. The singularities are regular. This is true because there is integral representation to start with. 

From Fact 2 it follwos that there is connection which is a rational function. From Fact 1 it follows that the denominator is a product of Landau polynomials. From Fact 3 it follows that these polynomials have to be in the first power.
\end{proof}

The purported significance of this theorem is that flat bundles over the complement of multicomponent divisors ( Landau divisor in our case) are in a sense more fundamental than Feynman integrals. In particular, they may be more amenable to the analysis. Let us formulate the following standard steps in the analysis of a flat bundle

{\bf S1:} Characterize the uniformization of multiple intersections of the Landau varieties The Landau polynomials have integer coefficeints. There are number fields ( extensions of rationals) associated to the diagram. The properties of these number fields must be reflected in the analytic properties of the hyperfunction $f$.

{\bf S2:} The calculation of the matrices $A$ involves several (multigraded) exact complexes, based on the standard $IBP$ and homogeneity relations ~\cite{GZK}. What is the homological algebra involved in this computation and what are the symmetries of the connection? The answer may be well known in general ~\cite{Hotta}.

{\bf S3:} There is canonical pull back of the flat bundle on each component of the singularity locus ( this is the "formalism of six functors", see ~\cite{Borel}). This restriction comes from asymptotic expansion of the hyperfunction in terms of Nilsson series, see ~\cite{Sturmfels}. Compute this restriction for each component of the Landau divisor. How do these restrictions match along intersections of components of the Landau divisor? (This is codimension 1 the Riemann-Hilbert data for the hyperfunction $f$ ~\cite{Sauloy}).

Note that these questions are valid for integrals of arbitrary polynomials , not just for Feynman integrands. 

In conclusion, we wish to formulate the following basic question about the general nature of a flat bundles over complements

{\bf Question:} Given a set of polynomals $L_1(z),...,L_k(z)$ with integer coefficients in the variables $(z_\mu)=(z_1,...,z_n)$ what is the structure of the space of polynomial matrices $A^{i_1,...,i_r}_{\mu,\beta,\beta'}(z)$ such that integrability conditions for the following system
\begin{equation}
\partial_\mu f_\beta(z)=\sum_{\beta'}\sum_r \sum_{i_1,...,i_r} \frac{A^{i_1,...,i_r}_{\mu,\beta,\beta'}(z)}{L_{i_1}(z)...L_{i_k}(z)}f_{\beta'}(z)
\end{equation}
are satisfied? The connection is assumed to be regular "at infinity" of the space $\{(z)\}$

It does not seem to the author that the answer is obvious. From the above, we know that at least for some polynomials $L_i$ this space is not empty. 

\bibliographystyle{amsplain}
\bibliography{references}

\end{document}